 \documentclass[11pt]{article}

 \pdfoutput=1

 \usepackage[polutonikogreek,english]{babel}

 \usepackage{textgreek} 
 
\usepackage{bigints}

\usepackage[round]{natbib} 

\usepackage{graphicx,siunitx,xspace}
\usepackage{amsmath}
\usepackage{amssymb}
\usepackage{hyperref}
\usepackage{enumitem}
\usepackage{color}
\usepackage{enumitem}
\usepackage{float}
\usepackage{amsfonts,bm}
\usepackage{epstopdf}
\usepackage[mathscr]{euscript}

  \usepackage[normalem]{ulem}
  \usepackage{textcomp}

\usepackage[utf8]{inputenc}
\usepackage[T1]{fontenc}
\usepackage{pifont} 

\usepackage{chemformula}

\usepackage{float}
\usepackage{ifthen}
\begin{document}


%

\begin{center}
{\bf \Large 
  A translation of the paper 
  ``\,\textbf{\emph{Presentation of some observations that}} \\ 
  \textbf{\emph{could be made to shed light on Meteorology\;}}'' \\  \vspace*{2mm}
  by Johann Heinrich \citet[][]{Lambert_Meteorology_p60_1771},} 
\begin{center}\vspace*{-1mm} 
 --------------------------------------------------- 
\end{center}\vspace*{-2mm} 
{\bf \Large 
from the (old-French language) paper: 
``\,\textbf{\emph{Expos\'e de quelques \\  \vspace*{0mm}
obfervations qu'on pourroit faire pour r\'epandre \\  \vspace*{0mm}
du jour fur la M\'et\'eorologie}}\,'' 
\textbf{\emph{in the revue  \\  \vspace*{0mm}
Memoires de l'Academie Royale des Sciences \\  \vspace*{1mm}
et Belles-Lettres de Berlin.}}}
\vspace*{-3mm} 
\begin{center}
 --------------------------------------------------- 
\end{center}
\vspace*{-3mm}
{ \large\color{black}
Translated by Dr. Hab. Pascal Marquet 
}
\\ \vspace*{2mm}
{\color{black}  \large Possible contact at: 
    pascalmarquet@yahoo.com}
    \vspace*{1mm}
    \\
{\color{black} 
    Web Google-sites:
    \url{https://sites.google.com/view/pascal-marquet}
    \\ ArXiv: 
    \url{https://arxiv.org/find/all/1/all:+AND+pascal+marquet/0/1/0/all/0/1}
    \\ Research-Gate:
    \url{https://www.researchgate.net/profile/Pascal-Marquet/research}
}
\\ \vspace*{-3mm}
\end{center}

\hspace*{65mm} Version-2 / \today

\begin{figure}[hbt]
\vspace*{0mm}
\centering \vspace*{0mm} 
-----------------------------------------------------------------------------------------------------------------
\vspace*{-1mm}
\includegraphics[width=0.5\linewidth]{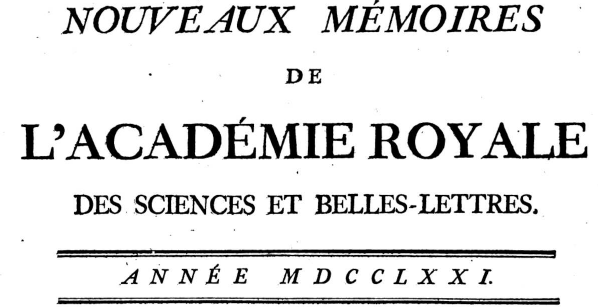} \\
\includegraphics[width=0.5\linewidth]{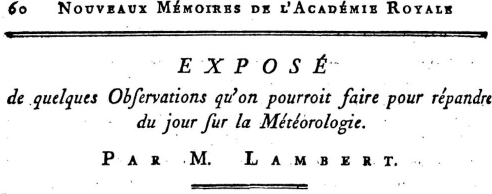}
\label{Figs_Lambert_Meteorology_1771}
-----------------------------------------------------------------------------------------------------------------
\end{figure}

\vspace*{2mm} 
\begin{center}
--------------------------------- \\
-------- Abstract ----------- \\
---------------------------------
\end{center}
\vspace*{-4mm}

\begin{quote}
{\centering
The Mulhouse mathematician Jean-Henri (or Johann Heinrich) Lambert (August 26 or 28, 1728; September 25, 1777) is well known for having devised the conformal conic projection in 1772, which is still used in some graphical outputs of our weather forecasting models, under the name ``{\it Lambert projection}.'' 
Less well known is that he also devised the idea of a minimum temperature value corresponding to $-270$°C in 1777, the year of his death and therefore 70 years before Lord Kelvin. 
But Johann Heinrich Lambert also published in 1771 an even lesser-known article, which is the subject of this publication. 
This article, written in Old French and published in a German journal, described in a rather striking manner the idea of a global observation network where the Earth would be divided into different zones where observers would record the same wind, temperature, pressure, and other current weather data at the same fixed times. The goal would then be to pool all this data to seek to understand how meteorological phenomena evolve in space and time, and to make meteorology a science on a par with astronomy.
}
\end{quote}




\vspace*{-2mm} 
\begin{center}
 --------------------------------------------------- 
\end{center}
\vspace*{-3mm}

\begin{center}
\includegraphics[width=0.34\linewidth]{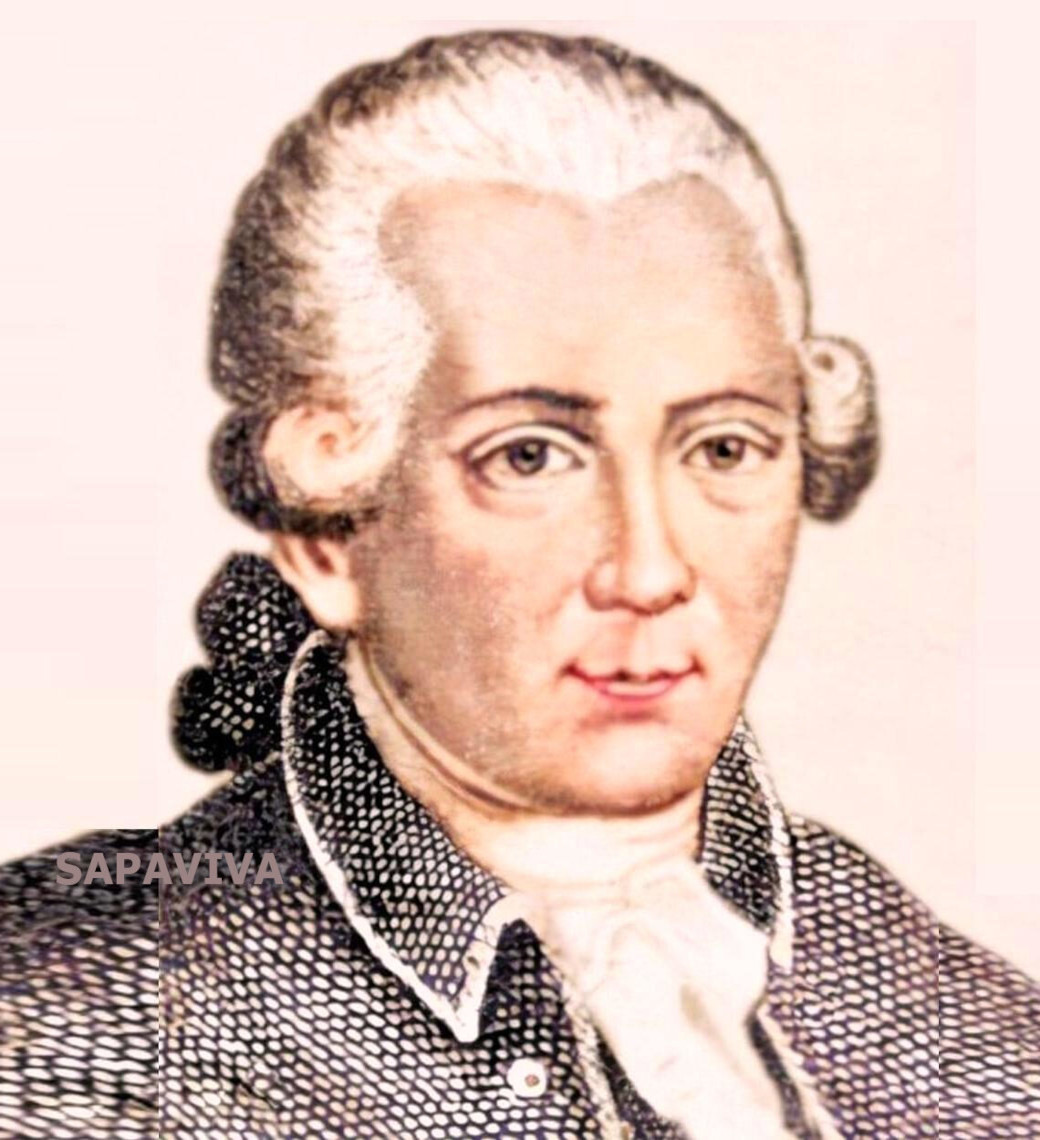}
\hspace*{15mm}
\includegraphics[width=0.28\linewidth]{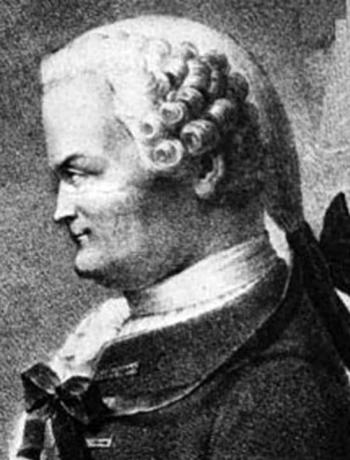} 
\\ 
Two photos of Jean-Henri (or Johann Heinrich) Lambert from 
\url{https://www.sapaviva.com/johann-heinrich-lambert/} and 
\url{https://fr.wikipedia.org/wiki/Jean-Henri_Lambert#/media/Fichier:JHLambert.jpg}
\end{center}


\section{The 1777-79 paper of J. H. Lambert: the zero of temperatures}

In this section, I have chosen to first discuss the remarkable contribution of Johann Heinrich Lambert concerning the definition of the zero of temperatures.

\begin{figure}[hbt]
\centering 
-----------------------------------------------------------------------------------------------------------------
\\ \vspace*{-2mm}
\caption{\it\small The thermometric scales published 
by Guillaume \citet[][p.53]{Amontons_Thermometre_Mesure_Fixe_1703_p50} 
(figure placed horizontally and with modifications to some legends for 
greater readability).
English translations of the French legends are: 
``Freezing of tallow'' (for ``Congélation du suif''); 
``Great heat of the 8th climate'' (for ``Grandes chaleurs du 8ème climat''); 
``Cellars at the Observatory'' (for ``Caves de l'Observatoire''); 
``Freezing of water'' (for ``Congélation de l'eau''); 
``Great cold of the 8th climate'' (for ``Grands froids du 8ème climat''); 
``degrees of heat'' (for ``degr\'es de chaleur''); 
``degrees of cold'' (for ``degr\'es de froid''); 
``New Thermometer'' (for ``Nouveau Thermometre''); 
``Old Thermometer'' (for ``Ancien Thermometre''); 
``very hot'' (for ``tr\`es chaud''); 
``hot'' (for ``chaud''); 
``temperate'' (for ``tempéré''); 
``cold'' (for ``froid''); 
``very cold'' (for ``tr\`es froid'').
}
\vspace*{1mm}
\includegraphics[width=0.7\linewidth]{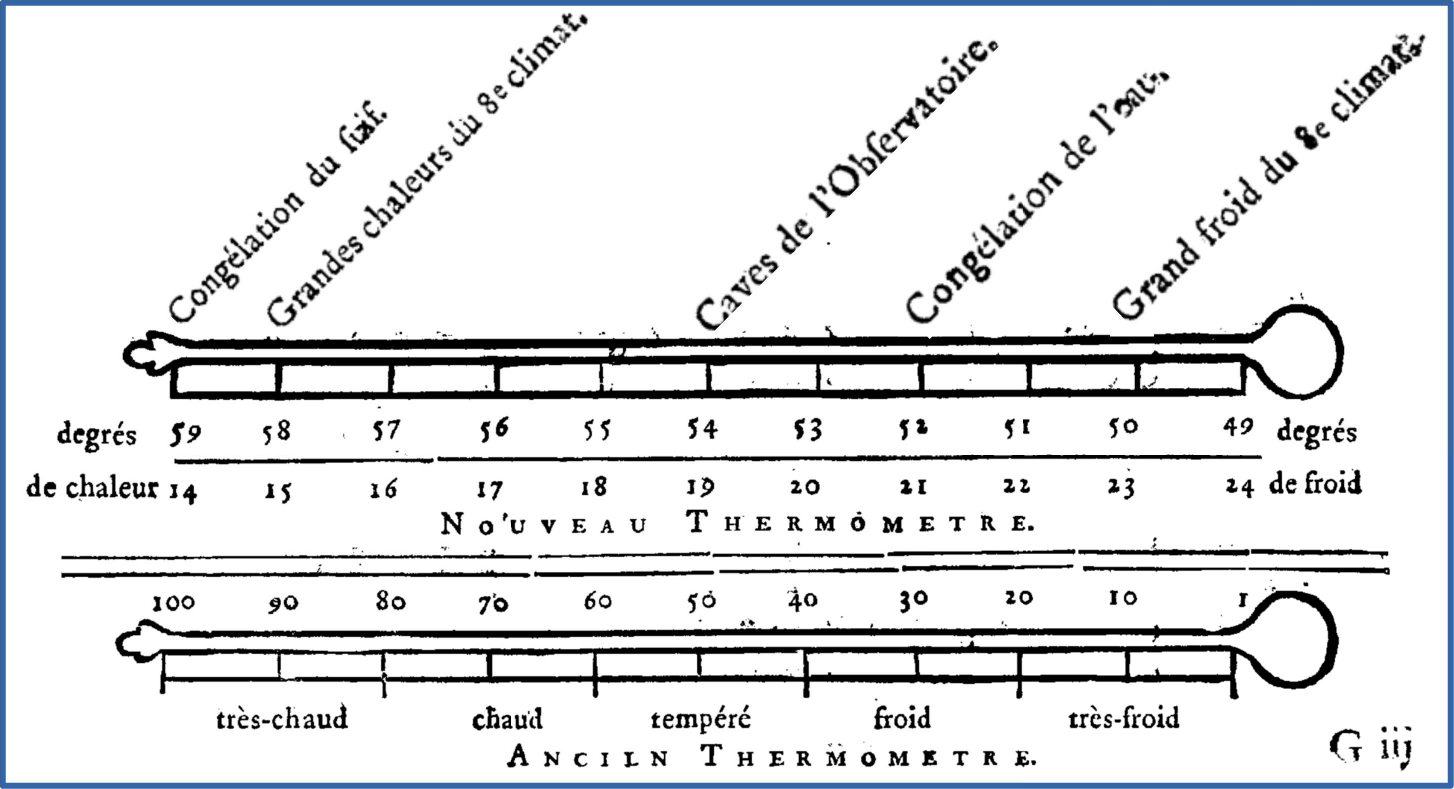}
\label{Fig_Amontons_1703}
-----------------------------------------------------------------------------------------------------------------
\end{figure}

It should first be mentioned that the French physicist, engineer and academic Guillaume Amontons (1663-1705) seems to be the first person to have imagined a value for the 
absolute zero of temperatures$\,$\footnote{$\:$This fact 
was brought to my attention by a video from Prof. Julien Bobroff in July, 2025.}. 
Indeed, in his article published two years before his death, 
Guillaume \citet{Amontons_Thermometre_Mesure_Fixe_1703_p50}  indicates on p.53 a new thermometric scale (see the Fig.~\ref{Fig_Amontons_1703}) and he writes on p.52-53 and in 
Old French$\,$\footnote{$\:$\label{footnote_old_french_a}In the Old French language used in the 1703 paper 
by Guillaume Amontons, 
the letter ``\,f\,'' replaces almost all the ``\,s\,'' and there was modified words like ``\,connoître\,'' instead of ``\,connaître\,'' and ``\,géle\,'' instead of ``\,gèle\,'', etc.}
the following sentence:

``{\it (...) since experience has shown us that if the heat of boiling water makes the spring of the air capable of supporting a load equal to that of $73$ inches of Mercury, the degree of heat which remains in the air when the water freezes is still great enough to make it support one equal to $51(1/2)$, which deserves very particular 
attention.\,}''$\,$\footnote{\label{footnote_inches_a}$\:$This refers to the ``\,French inch\,'' of the old regime, which was equal to  $2.707$~cm at the time (and therefore a little larger than the English Inch, which was already equal to $2.54$~cm). This French inch was divided into $12$ lines of $2.256$~mm, with a French foot being equal to  $12$~inches or $32.484$~cm, which corresponds to $1/3.078$ meter and therefore to approximately one third of a meter.}

We can deduce from Guillaume Amontons' sentence published in 1703 that the zero of the temperatures (where all the $51.5$~inches remaining at the freezing temperature of the water would be consumed) must be equal to  
$-100 \times 51.5 / (73-51.5)$, or approximately $-240$°C. This value is of the same order of magnitude as that of $-273$°C considered 145 years later by William \citet{Thomson_1848}, the future Lord Kelvin, who gave his name to the modern scale of absolute temperatures, which today is set at $-273.15$~C.

The ``\,absolute\,'' aspect, which Guillaume Amontons should deserve special attention, lay in his wish not to assign to the freezing temperature of water an arbitrary value such as zero in the Celsius, Newton or R\'eaumur scales, nor even to choose other arbitrary values for other temperatures (such as $100$ degrees for the boiling temperature of water in the Celsius scale, or $0$ and $100$ degrees Faherenheit for other temperatures). 
On the contrary, for Amontons, it was a matter of fully trusting the height of the air column used to evaluate the temperature, a bit like the $76$~cm height of the mercury column which corresponds to a standard atmosphere, with the only unknown being the scale chosen to measure this length.

The 1911 Encyclopaedia Britannica entry for the article ``\,Cold\,'' 
(\url{https://en.wikisource.org/wiki/1911_Encyclop\%C3\%A6dia_Britannica/Cold}) 
reads as follows (after some historical checking and corrections/additions): 
``{\it Values of this order for the absolute zero were not, however, universally accepted about this period. 
Laplace} (in fact ``\,de la Place\,'') {\it and Lavoisier, for instance, in their treatise on heat \citep[][p.384]{Lavoisier_Laplace_1780}, arrived at values ranging from $1500$° to $3000$° below the freezing-point of water, and thought that in any case it must be at least $600$° below} (p.385), {\it while John \citet{Dalton_1808} in his ``\,Chemical Philosophy\,''} (section 6 ``\,On the natural zero of temperature or absolute privation of heat\,'') {\it gave ten calculations of this value, and finally adopted} (p.97) $-3350$°C (in fact $-6000$°F) {\it as the natural zero of temperature.}'' 

\begin{figure}[hbt]
-----------------------------------------------------------------------------------------------------------------
\vspace*{-2mm}
\caption{\it
{\bf On the left:} 
The ``{\it\,Table of observed degrees of heat\,}'' 
published in the §-508 (p.276-277) in the book 
by \citet[][]{Lambert_Pyrometrie_1779}, 
with Fahrenheit temperatures from $-62$°F to $1600$°F.
{\bf On the right:}
Two diagrams of the  
``{\it\,degrees of the air thermometer\,}'' 
plotted from the same Table 
by \citet[][]{Lambert_Pyrometrie_1779}, 
but against the Celcius temperatures 
from $-320$°C to $950$°C (Top), and a zoom 
for temperatures from $-320$°C to $140$°C (bottom).
}
\centering \vspace*{2mm}
\hspace*{-12mm}
\includegraphics[trim=10 105 20 35,clip,width=1.14\linewidth]{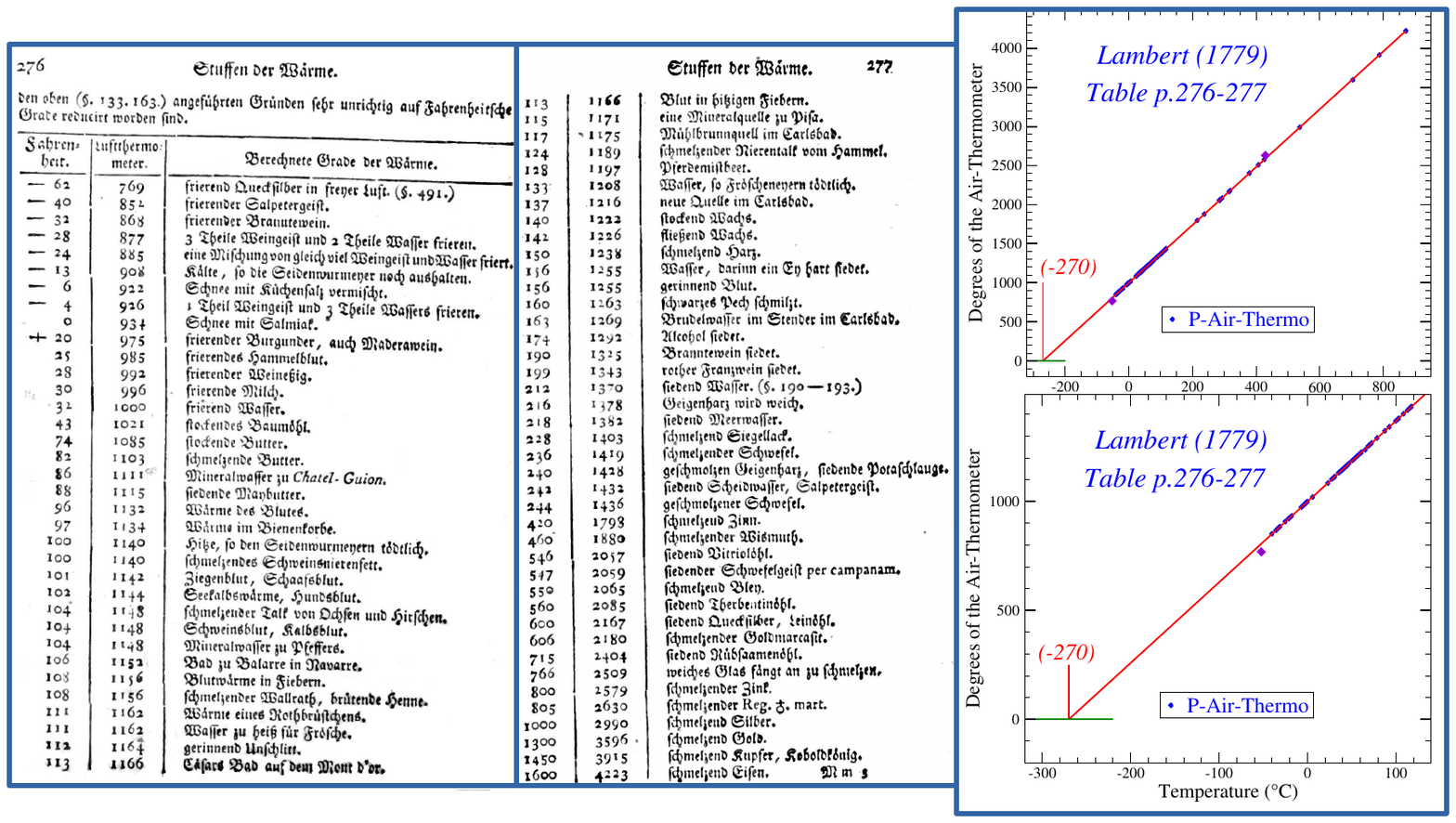}
\label{Figs_Lambert_1779}
-----------------------------------------------------------------------------------------------------------------
\end{figure}

Far from the very low values suggested by Lavoisier, Laplace, or Dalton, who disagreed with the more moderate value $-240$°C of Amontons, in a work published two years after his death  
Johann Heinrich \citep[][p.273-277]{Lambert_Pyrometrie_1779} 
published a table of \;``\,degrees of the air thermometer\,'' 
measured as a function of temperatures expressed in degrees Fahrenheit.
From this table, I was able to plot the Figs.~\ref{Figs_Lambert_1779} for this article, where it appears that most of the points (except two) are aligned along a straight line that can be extended to a minimum temperature of approximately $-270$°C, where the degrees of the air thermometer would cancel out. 
Only the points $-52.2$°C ($-62$°F) and $429.4$°C ($805$°F) do not exactly align along the straight lines.
These Figs.~\ref{Figs_Lambert_1779} correspond to the Figure~1 in \citet{Marquet19LMa}, where I have outlined the principle of calculating absolute temperatures in connection with the third law of thermodynamics and the hypothesis of the unattainability of absolute zero \citep[see the Figure~3 in][]{Marquet19LMa}, which already according to the words of Thomson-Kelvin in 1848: ``{\it \,is a point which cannot be reached at any finite temperature, however low.}\,''

We can therefore see that J. H. Lambert suggested in 1777-79 a value for the zero of temperatures which was comparable to that of Amontons and using the same intrepid hypothesis of an extrapolation which would be licit towards the zero of the degrees of the air thermometer (i.e.~of the height of a certain measured column). 
J. H. Lambert also explained (p.29) that this zero would correspond to an air at ``\,absolute cold\,'' (\,``{\it absoluten Kälte}\,'' in German) where the air would collapse on itself (\,``{\it fälle}\,'' in German) and would condense tightly towards a stage where its ``{\it particles}\,'' (\,``{\it Teilchen}\,'' in German, but in the sense of the ``{\it Calorique}\,'' of the time) would be in direct contact with each other.

Of course, this Caloric hypothesis would not survive future visions of thermodynamics with the subsequent appearance of the atomistic vision of matter, but in any case, the value of $-270$°C of J. H. Lambert in 1777-79 deviated greatly from the unrealistic values of Lavoisier, Laplace and Dalton, while being very close to that of $-273$°C which would be retained from 1848 by Lord Kelvin, almost $70$ years later.

It should be noted that the common point between the studies of Amontons, Lambert and Kelvin is the use of the old ``{\it constant volume air thermometers}\,'' for which the pressure varies linearly with the temperature ($p=a*t-b$) and with the theoretical possibility of extending the lines to the absolute zero of the temperatures $t=b/a$ for which $p=0$. 
The study of other types of thermometers, such as those with liquid columns of alcohol or mercury, would not have allowed this theoretical advance, insofar as the height of these columns does not tend towards $0$ at absolute zero of temperatures. 
This attention of the 
mathematician$\,$\footnote{$\:$The Wikipedia page recalls that J. H. Lambert, in addition to his work on conformal transformations and projections on the sphere, was the first to demonstrate the irrationality of the number $\pi$ in 1767; introduced hyperbolic functions; introduced the $W$ function (known as Lambert's function, in the context of solving equations such as $x + e^x = a$); with work on ruler and compass plots which led him to discuss Euclid's famous fifth postulate on parallels (1786).}
J. H. Lambert for experimental measurements could explain his motivation for the global observation system on the Earth described in the following section.


\vspace*{-2mm} 
\begin{center}
 --------------------------------------------------- 
\end{center}
\vspace*{-3mm}

\section{The paper by J. H.~\citet{Lambert_Meteorology_p60_1771}: a global system of observations}

In this second section I have chosen to reproduce verbatim 
the text by J. H.~\citet{Lambert_Meteorology_p60_1771}, 
but with a translation from a transposition 
from the old into the modern French language, 
in order to facilitate reading of the text 
in 2025$\,$\footnote{$\:$\label{footnote_old_french_b}Just 
like in the footnote \ref{footnote_old_french_a}, in the Old 
French language used in the 1771 paper by 
J. H.~\citet{Lambert_Meteorology_p60_1771}   
the letter ``\,f\,'' replaces almost all the ``\,s\,'' 
and there was modified words like:  
``\,faudroit\,'' instead of ``\,faudrait\,''; 
``\,loix\,'' instead of ``\,lois\,'';
``\,{\&}\,'' instead of ``\,et\,''; 
``\,mouvemens\,'' instead of ``\,mouvements\,''; etc.}.
I have also added several footnotes in order to provide additional information, 
but to leave the original text unaltered. 

\vspace*{0mm} 
\begin{center}
-------------------------------------------------------------------- \\
-------- The English translated 1771 paper ----------- \\
-------------------------------------------------------------------- 
\end{center}
\vspace*{-4mm}


It seems to me that to make Meteorology more scientific than it is, we should imitate the Astronomers who, without considering all the details, begin by establishing general laws and average movements.
In this way they are able to take anomalies into account, and subject them to laws.
In this way they finally managed to predict phenomena with an accuracy that inspires, even to the most ignorant, respect for Astronomy.

What is the same about meteorology? It is very certain that it has general laws, and that a large number of periodic phenomena enter into it.
But these latter can hardly yet be guessed.
The observations made so far, among which there is no connection, are of little importance. 
Many changes, especially in the gravity of air, have their causes in distant countries. 
The important point, therefore, would be to understand their 
sequence$\,$\footnote{$\:$It should be remembered that J. H. Lambert's 1771 article preceded by almost 50 years the work of Henri Navier, who in 1822 introduced the notion of viscosity into Euler's equations, and by more than 70 years the work of George Gabriel Stokes, who in 1845 gave its definitive form to the equation for the conservation of momentum. This means that J. H. Lambert had reason in 1771 to hope to discover unknown laws of variation of meteorological conditions through the examination of observed data alone, and without the support of the theoretical laws of Navier-Stokes, which were not known to him. 
Similarly, the 1771 paper by J. H. Lambert's preceded by 53 years the book by 
Carnot (1824), which was at the basis of modern thermodynamics, 
not to mention that we had to wait for the later work of 
Clapeyron (from 1834 to 1847), 
Regnault (1847), 
Thomson-Kelvin (from 1848 to 1879), 
Rankine (from 1852 to 1855),
Kirchhoff (from 1858 to 1894),
Joules (from 1844 to 1857), 
Clausius (from 1850 to 1867), 
Gibbs (from 1875 to 1878), 
Helmholtz (from 1882 to 1888), etc, 
to really understand the thermodynamic processes which are 
at the basis of the evolution of the moist atmosphere.}. 
To achieve this, neither great expense nor much preparation will be required.
And if considerable sums have been spent on some astronomical observations, meteorology, which is of such close interest to the whole human race, would also deserve to be devoted to it. 
This is what it all comes down to.

It is known that great barometric variations occur at the same time, or almost at the same time, over a large area of country, and I am not mistaken in saying that the course of the barometer is, with a few slight differences, the same in Russia as in Portugal.
However, these analogous variations extend more in longitude than in latitude. For if, for example, the barometer under the Pole varies by 3 inches ($8$~cm), it only varies by 2 inches ($5.5$~cm) in Paris, and we know that between the Tropics its variations do not go beyond 3 or 4 lines 
($7$ to $9$~mm)$\,$\footnote{$\:$Like in the footnote~\ref{footnote_inches_a}, the old French inches (of $2.707$~cm) was separated into $12$ lines (of $2.256$~mm), with one French foot equal to $12$ French inches or $32.484$~cm, which is about $1/3.078$ of a meter.}.

\begin{figure}[hbt]
\centering 
-----------------------------------------------------------------------------------------------------------------
\vspace*{-2mm}
\caption{\it\small Two examples of convex regular icosahedra formed from 20 triangular faces and approaching the sphere (one of the Platonic solids): on the left from  
\url{https://www.aquaportail.com/dictionnaire/definition/11550/icosaedre}; 
and on the right from \url{https://fr-academic.com/dic.nsf/frwiki/833227}}
\centering \vspace*{0mm}
\includegraphics[width=0.29\linewidth]{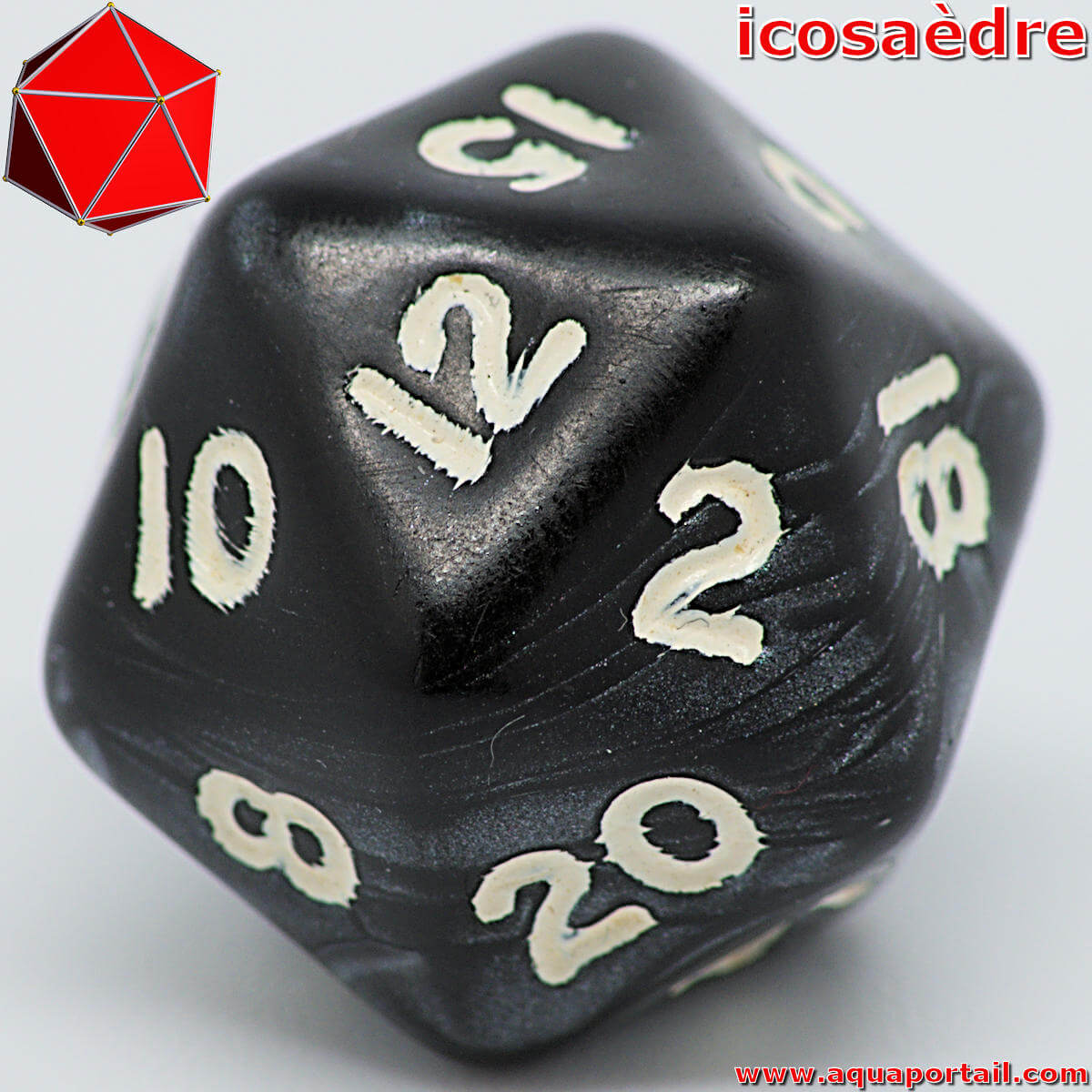} \hspace*{20mm}
\includegraphics[width=0.27\linewidth]{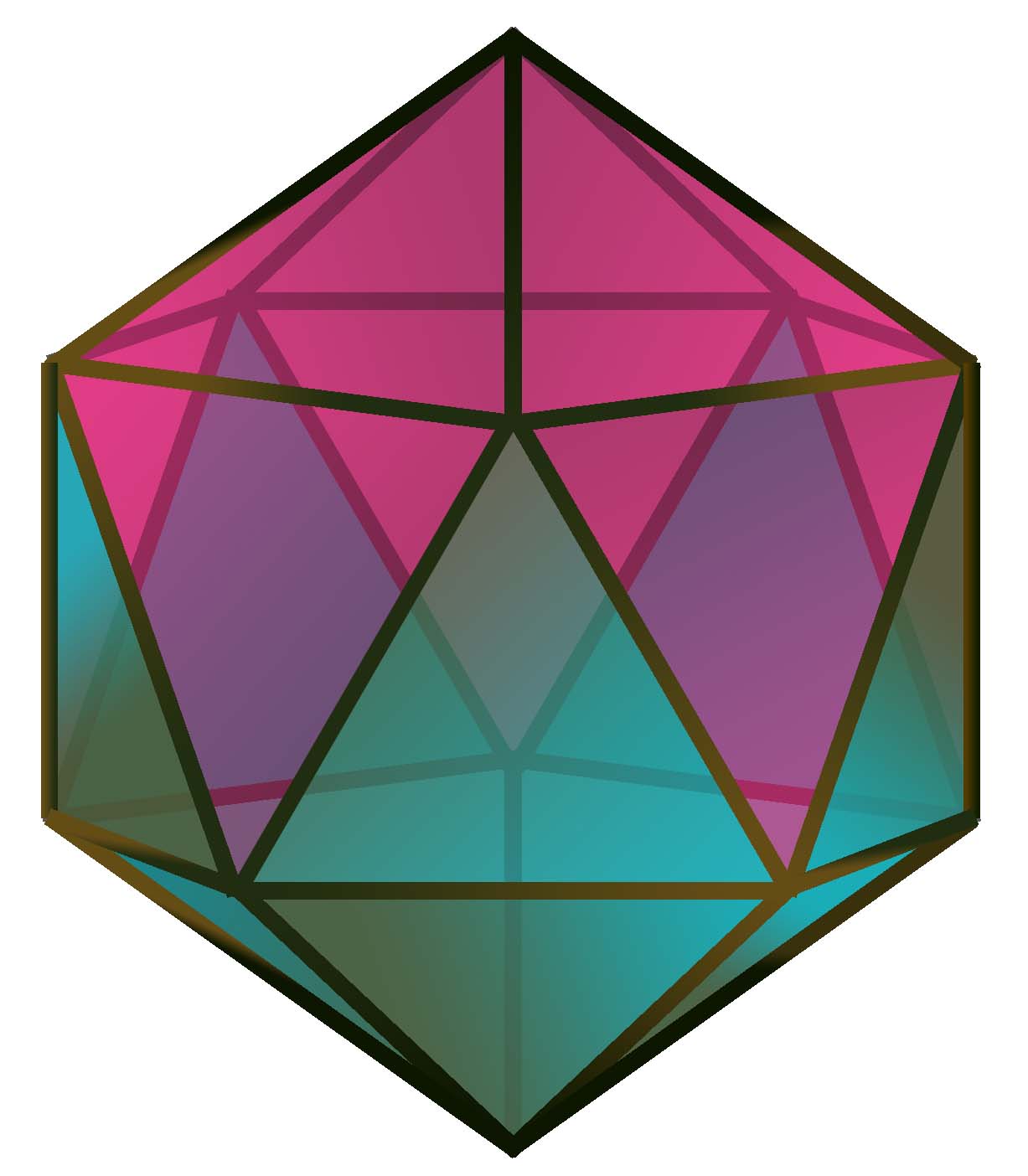}
-----------------------------------------------------------------------------------------------------------------
\label{Figs_Isocaedres}
\end{figure}

This is what we observed, but we do not know how it happens.
Perhaps the shrinking of Zones and Climates is contributing to this.
But observations made according to a certain method will establish them for us infinitely better and in more detail.
If it were absolutely feasible, I would divide the surface of the Earth into 20 equal triangles, therefore in the shape of an isocahedron 
(see the Figs.~\ref{Figs_Isocaedres}), 
and at the center of each of these triangles meteorological observations would be made.
In addition to these 20 Observers, 12 could be placed at the points of intersection of these triangles; and thus with 32 Observers, one would be able to keep a record of all the revolutions of the Atmosphere which would be subject to some general law.
The whole weight of the atmosphere, its equilibrium, and the way in which it changes and re-establishes itself could be taken into account. These observers would be separated from each other by 37 to 41 degrees; and this distance would not be too great, seeing that the variations of the barometer, which depend on some general law, extend, if not further, at least just as far.

\vspace*{2mm}

\begin{figure}[hbt]
\vspace*{-4mm}
\centering 
-----------------------------------------------------------------------------------------------------------------
\vspace*{-2mm}
\caption{\it\small 
The figure inserted between the pages 64 and 65 of
\citet[][]{Lambert_Meteorology_p60_1771} 
(from: \url{https://pictures.abebooks.com/inventory/31862847446.jpg}): 
{\bf Top:} the original version (with an old, somewhat crude background map); 
{\bf Bottom:} a version where I have plotted the same observation points 
(in red) on a modern global background map (without points on Australia or 
the west of the current USA).}
\centering 
\vspace*{2mm}
\includegraphics[width=0.55\linewidth]{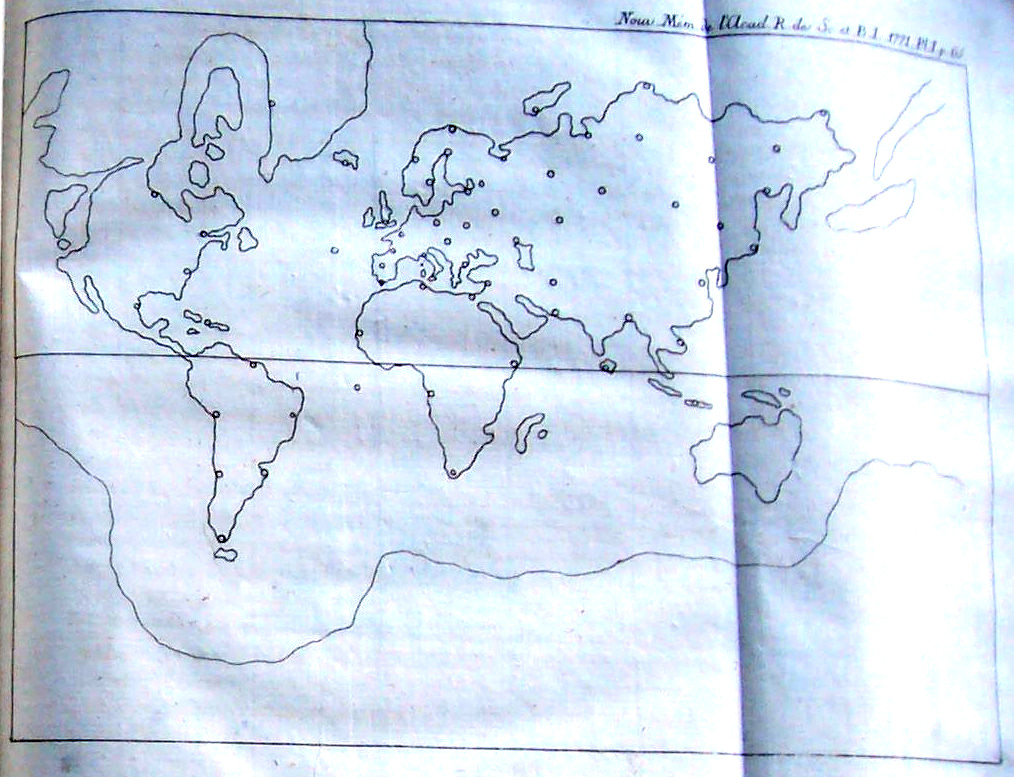}
\includegraphics[width=0.65\linewidth]{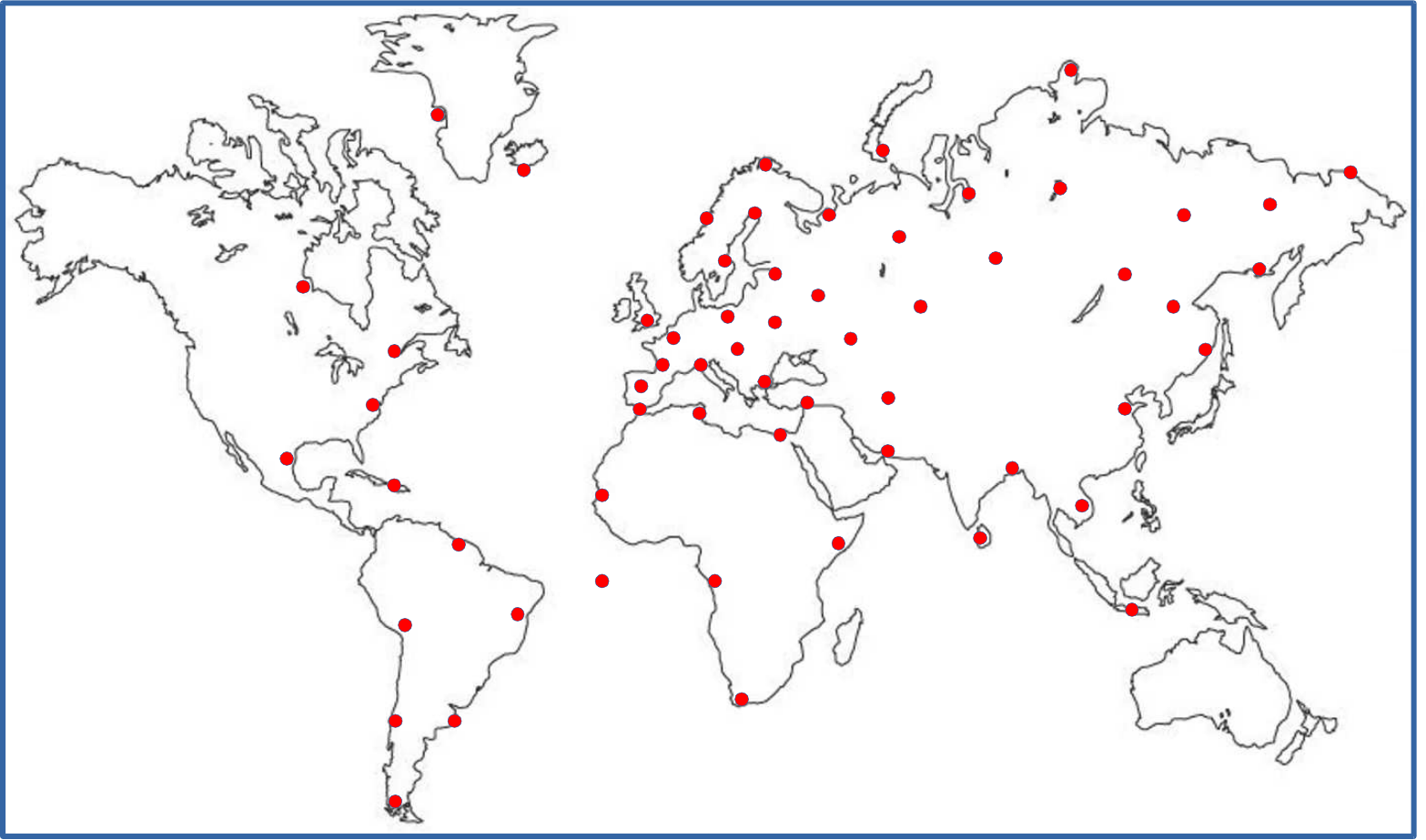}
-----------------------------------------------------------------------------------------------------------------
\label{Lambert_1771_Figure}
\end{figure}

This project could be carried out, if one were willing to bear the expense. 
Nothing is easier than to determine these triangles on the globe. 
But one can also deviate more or less from this absolute regularity; and then the expense is reduced to an unimportant thing. 
Instead of placing observers at regular, but perhaps inaccessible, inconvenient, and dangerous points, it would be enough to choose places where trading nations have established colonies, where there are missionaries, in a word, where there are people who know how to count and write; for that is all that is needed for meteorological 
observations$\,$\footnote{$\:$The training of meteorological observers can undoubtedly be as succinct as what J. H. Lambert describes here. But the training currently received, for example, in the Schools of Meteorology (like the ENM in France) is much broader, ranging from courses on various measuring instruments to courses in thermodynamics and atmospheric kinematics and dynamics. But J. H. Lambert could not have anticipated this in 1771.}.

In passing over the globe I looked for the places, or at least the surroundings of the places which would form, or almost form, triangles such as I have just described (see the Figure~\ref{Lambert_1771_Figure}). 
It is true that more than one Nation should be interested in arranging things so that in such and such a year the meteorological observations would begin at the same time, in accordance with the Instruction which would be drawn up in this regard.

I believe, however, that the English nation could begin to set the tone, and there is no doubt that other trading nations would contribute to complete the whole system. 
For this purpose, it would first be sufficient to publish in the Gazettes the resolution that has been taken. 
And the Royal Society of Sciences of London, after having drawn up the Instruction and the plan for these observation forces, would publish this plan, and send it to the Academies established in other trading countries.
And as in Europe there are many amateurs of meteorological observations, I believe that several of them would conform to it without difficulty, since at least for Europe the simultaneous and successive state of the Atmosphere would be known to the point of finding its more particular laws.

I do not know whether the Pilots, when recording the winds in their logs and even determining the strength of the wind by the speed of the ship and by the state of the sails and the drift, also record the weather; but nothing could be easier.
And although a ship at sea is a mobile observatory, the large number of ships which are always on a voyage would not fail to make known the state of the Atmosphere in these intermediate regions where observers could not be 
found$\,$\footnote{$\:$J. H. Lambert's 1771 plea for possible meteorological measurements on ships was highly innovative and prophetic, anticipating future (stationary) Weather Ships and SHIP messages sent by ships of all nations traveling the oceans.}.

To avoid making these observations difficult, until the results are seen, a barometer and a thermometer may suffice at first.
The scales will be either corresponding or at least 
intelligible$\,$\footnote{$\:$J. H. Lambert refers here to the multitude of thermometers available in 1771 and the need to be familiar with the formulas for transcribing all observations onto the same given scale, to facilitate the comparison and study of all these observations, with for example those of: Celsius, Fahrenheit, Réaumur, La Hire, Amontons, de l’Isle, Newton, du Crest, Poleni, Mariotte, Benart, Fowler, Hales, ... 
See the table VIII at the end of the book by \citet[][]{Lambert_Pyrometrie_1779}, with the Figures and tables available at:  
\url{https://docnum.unistra.fr/digital/collection/coll12/id/61486}.}. 
The locations where observations will be made will be, as far as possible, at sea level, in order to have the weight of the entire atmosphere everywhere. The weather will be noted, as well as the winds and their different degrees of strength. 
As for the time, and to take into account the difference of the Meridians, one could take the noon of London. In this way one will observe the evening in the East Indies, and the morning in the West Indies, in order to have observations made at 
the same time$\,$\footnote{$\:$In 1771, J. H. Lambert could not, of course, have known Universal Time (UT), which, from 1972 onwards, became the international time scale based on the Earth's rotation. But J. H. Lambert could not have known Greenwich Mean Time (GMT) in 1771 either, which was not adopted until more than a hundred years later, in 1884 at the international conference in Washington. Moreover, J. H. Lambert's 1771 article was published 120 years before the time was fixed in France and Algeria at Paris Mean Time (in 1891) to meet the needs of the rise of rail transport. J. H. Lambert's 1771 plea for an international definition of time was therefore highly innovative and prophetic.}.

We could have copies of the registers sent to us every year.

\vspace*{-2mm} 
\begin{center}
 --------------------------------------------------- 
\end{center}
\vspace*{-3mm}

Here is how I believe they could be arranged.


They can be divided into four parts.

The first part will look at barometric observations. Each month will be given a folio page. This page will have as many columns as there are places where observations were made. And next to each day, the height of the barometer will be marked, which in each column corresponds to that day.
The columns will follow the order of the longitude of the places. As for the barometric heights, it will be sufficient to express them in lines (of $2.256$~mm) and decimal parts of lines. One could even limit oneself to marking by how much they exceed the $300$ duodecimal lines of the foot; this will be to narrow the columns and to make comparison 
easier$\,$\footnote{$\:$Here we recognize in J. H. Lambert's remarks the concern for digital compaction still in use in data transmission or storage systems: coding of deviations from reference values such as for average temperatures and pressures at a given altitude.}.

The second part will be arranged in the same way with respect to the thermometer, which can be that of Fahrenheit, because its zero is low enough that the difference between hot ({\it i.e. positive Fahrenheit values $>-17.8$°C}) and cold ({\it i.e. negative Fahrenheit values}) need only be marked 
very rarely$\,$\footnote{$\:$It may be recalled that the $0$ and $100$ of the Fahrenheit scale proposed in 1724 corresponded to somewhat arbitrary extreme values: the lowest ($0$°F$\:=-17.78$°C) was perhaps that measured during the winter of 1708/1709 in his native Danzig (although later associated with the solidification temperature of a eutectic mixture of ammonium chloride and water); the highest being that of the human body under the armpits or in the mouth. J. H. Lambert's remark must therefore be understood as the fact that for most observations we must have ``\,hot\,'' (positive) values above $0$°F$\:=-17.78$°C, and few ``\,cold\,'' values (negative in Fahrenheit).}.

The third part will be arranged in the same way with respect to the winds. One, two, three points would be added to the letters S, N, W, O, to mark one, two, three degrees, etc. of force. But with regard to winds which rise or cease suddenly, it would be good to mark there also the time when they began or ended.

\begin{figure}[hbt]
\vspace*{0mm}
\centering 
-----------------------------------------------------------------------------------------------------------------
\vspace*{-2mm}
\caption{{\it 
{\bf Left (a,b):} The symbols for the weather conditions shown on page 63 in the 1771 article by J. H. \citet[][]{Lambert_Meteorology_p60_1771}. 
Here we recognize the beginnings of the symbols used to represent the present weather conditions or clouds on operational weather charts (such as the ``\,Norwegian\,'' charts colored in red, blue, purple, etc.). 
{\bf Lambert's French terms} were: 
``\,nuées\,'' for clouds; 
``\,pluie\,'' for rain; 
``\,neige\,'' for snow; 
``\,brouillard\,'' for fog; 
``\,tonnere/orage\,'' for lightning/thunderstorm.
{\bf Right (c):} the modern symbols of the present weather conditions for the French ``\,temps significatif\,'' (or Significant Weather) Aeronautical Charts (TEMSI), with almost identical symbols for rain and snow, and similar for thunderstorm.}}
\vspace*{1mm}
\hspace*{0mm}
\includegraphics[width=0.8\linewidth]{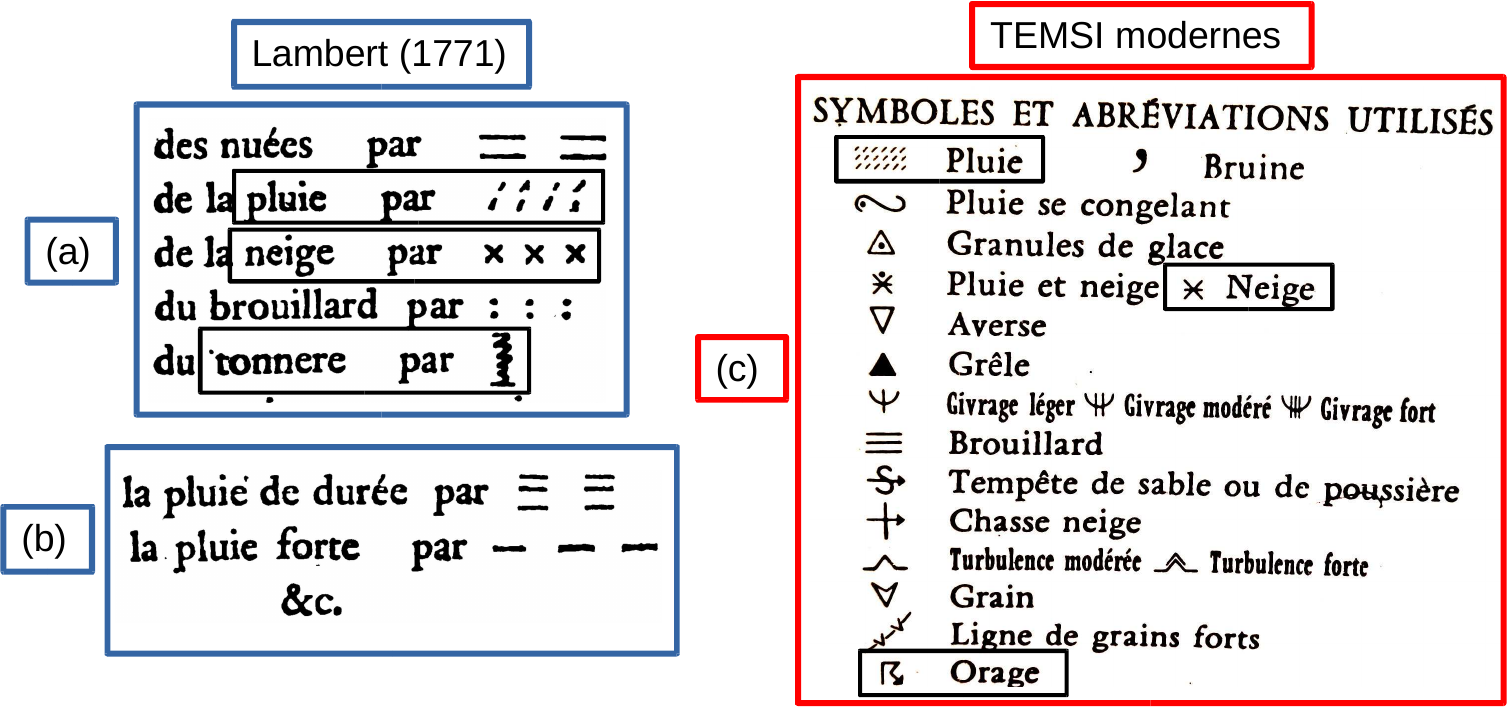}
\label{Fig_Lambert_Meteorology_1771_p63}
-----------------------------------------------------------------------------------------------------------------
\end{figure}

Finally, the fourth part, which will be arranged in the same way, will offer the visible and sensible state of the air. 
As in each column there is a square cell for each day, this cell can be left blank for serene days. 
As for the other days, one could mark  
{\it the symbols shown in the part (a) of the 
Fig.~\ref{Fig_Lambert_Meteorology_1771_p63}}, 
and in this way one could also, by doubling these signs, 
mark the degree of strength and duration, with for example 
{\it those in the part (b) of the 
Fig.~\ref{Fig_Lambert_Meteorology_1771_p63}}.

It is clear that in this way the simultaneous and successive state of the air will be obvious, and that it will be seen as if at a glance.

In the comparison that can then be made between these observations, small changes will hardly come into play, since they ordinarily depend on accidental causes, which do not extend their effect very far.
But the great variations will be of greater consequence, in that they extend further and have effects of a longer duration. It is especially to these (great variations) that we must have regard, in order to discover the general laws of the simultaneous and successive sequence of time, and in what manner its changes influence from one country to another, from one climate to another, and finally from one hemisphere to another.

Thus, for example, in Europe there are sometimes entire months which are fine or rainy. How far does this extend and from which country does the change then come? This is where we may have occasion to find general laws. For example, can the causes of these changes be found on the ocean, or do they always come from some dry land?

It is also for this reason that in Europe the North-East wind brings good weather, while the opposite South-West wind brings rain. Where does this begin and end? And why is it that sometimes, and especially in certain seasons, this does not happen, and that when the North or East wind brings rain, it is usually a very widespread and very long-lasting rain?

Moreover, despite all the small irregularities, there is something periodic in the variations of the weather, even if it is only the daily and annual variations of cold and heat. 
This must lead to other more or less periodic phenomena. 
Now, by considering them in this way, on a large scale and relative to the whole Atmosphere, there is no doubt that we will find general laws and different causes of the anomalies which 
are considerable$\,$\footnote{$\:$We can easily find in these sentences of J. H. Lambert a prophecy of the discovery of baroclinic disturbances with their warm and cold fronts and occlusions, but also of cyclones and even squall lines: it is undoubtedly these large-scale ``\,periodic\,'' phenomena that J. H. Lambert wished to see highlighted through the study of the observation network that he imagined in 1771.}.
There are still a thousand questions which can be clarified by means of these kinds of registers. We will see, for example, which districts are covered by clouds, what difference there is according to the seasons, what path the clouds take, how far the strong and variable winds extend, where they originate, how they counterbalance each other, where this happens, what relation they have to the variation of the weight of the air, why is it that, while during the equinox a part of the atmosphere passes the Equator, this hardly causes the barometer to vary at all, whereas under the Pole it can vary by 3 inches? etc.

If, at least in the main locations, we could add to the observations indicated above those of the quantity of rain, the evaporation of the hygrometer, the declination and inclination of the compass, the system would become more complete.
As for hygrometers, I have already shown in my Essay on Hygrometry (Mém. de l'Acad. 1769), and I will show it again on other occasions, that these instruments can be made to correspond. Very recent observations, made over a whole year in Silesia by the Baron de Felbiger, very worthy Prelate of the Abbey of Sagan, and in Berlin by myself, have shown me that the variations in humidity are very similar and very often equal in these two locations.

\vspace*{0mm} 
\begin{center}
----------------------------------------------------------------------------- \\
-------- End of the English translated 1771 paper ----------- \\
----------------------------------------------------------------------------- 
\end{center}
\vspace*{-3mm}



\vspace*{-3mm} 
\section{Conclusion}

According to Wikipedia (\url{https://en.wikipedia.org/wiki/Timeline_of_meteorology}), 
Johann Heinrich Lambert's 1771 article preceded by 78 years 
the observation network established in 1849 across the United States by the Smithsonian Institution, with 150 observers via telegraph, under the leadership of Joseph Henry.
Johann Heinrich Lambert's 1771 article also preceded by 82 years 
the first International Meteorological Conference held in 1853 in Brussels at the initiative of Matthew Fontaine Maury, U.S. Navy, with the recommendation of a standard observing times, methods of observation and logging format for weather reports from ships at sea (just like anticipated and predicted by Johann Heinrich Lambert in 1771).

In  France, Johann Heinrich Lambert's 1771 article preceded by 83 years 
the maritime disaster linked to the Crimean storm of November 14, 1854, which motivated Le Verrier's request to Napoleon III to create a network of meteorological observatories on French territory (created in 1856). 
Of course, J. H. Lambert's network (20 + 12 = 32 observation stations around the globe) was very minimal compared to the 24 and then 59 stations created throughout Europe in 1865, partly linked by telegraph. 
But J. H. Lambert's vision was nevertheless prophetic in this matter, and I hope that the present translated article could usefully make its value known and appreciated.

\vspace*{4mm}

\newpage

\bibliographystyle{ametsoc2014}
\bibliography{Book_FAQ_Thetas_arXiv}

\newpage 
\noindent 
\hspace*{50mm}
-----------------------------------------------------------------------\\  
\hspace*{53mm} -------- The old-French written 1771 paper -----------\\ 
\hspace*{50mm}
-----------------------------------------------------------------------
\vspace*{4mm} 

\begin{figure}[hbt]
\vspace*{-8mm}
\centering 
\includegraphics[width=0.9\linewidth,trim=0 38mm 0 0,clip=true]{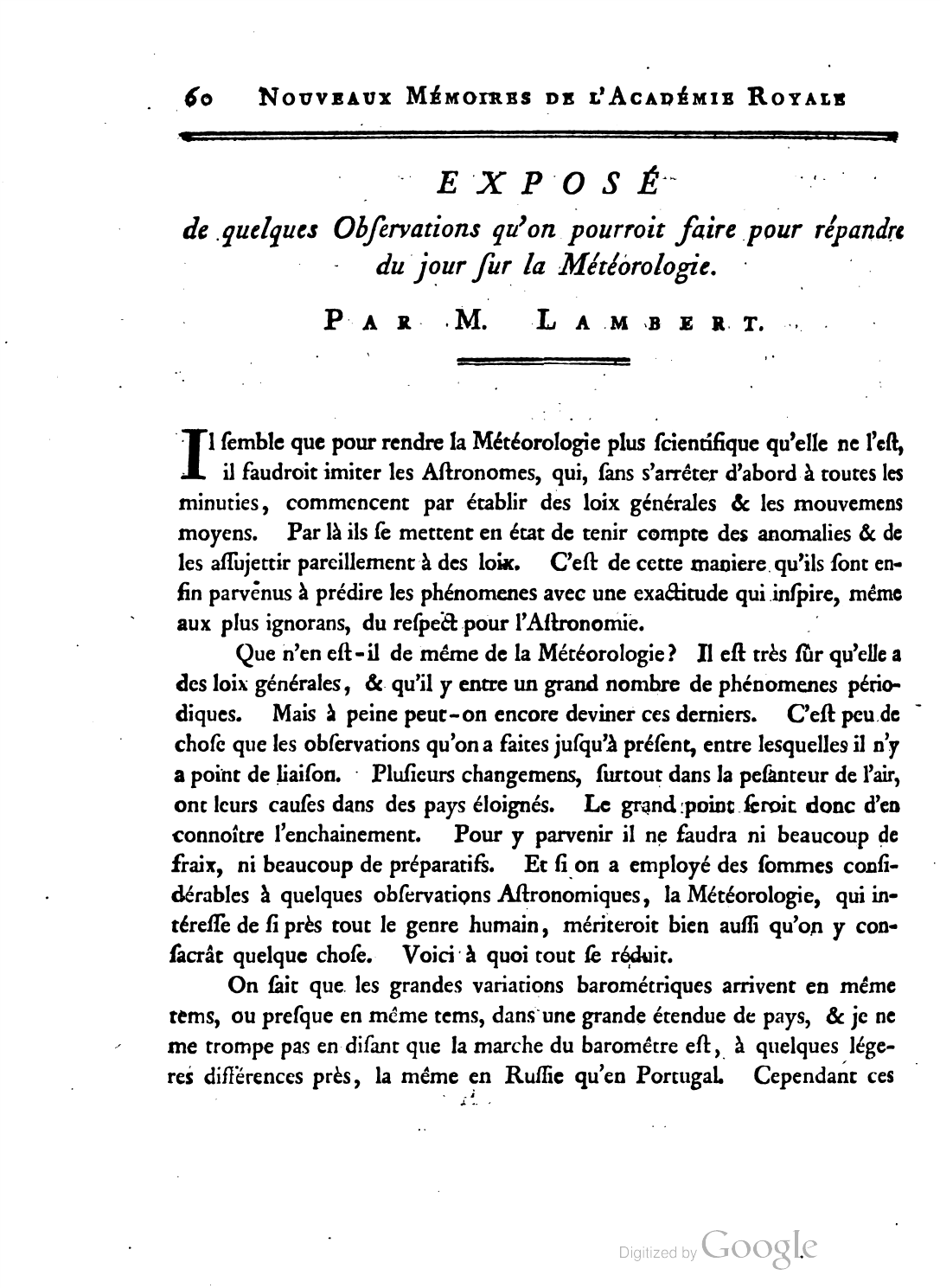}
\label{Lambert_1771_p60}
\end{figure}
\clearpage

\begin{figure}[hbt]
\vspace*{-4mm}
\centering 
\includegraphics[width=0.9\linewidth,trim=0 36mm 0 0,clip=true]{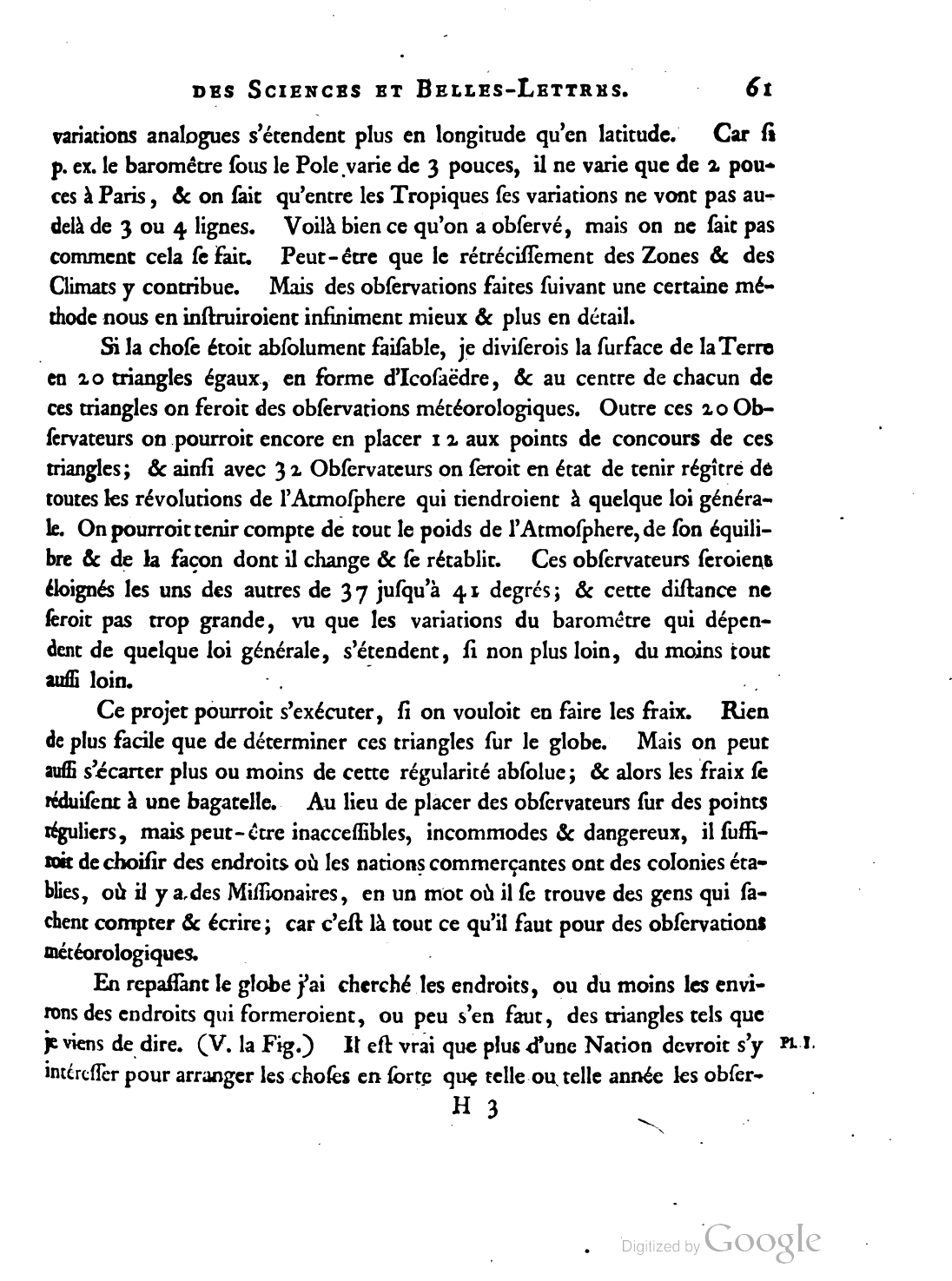}
\label{Lambert_1771_p61}
\end{figure}
\clearpage

\begin{figure}[hbt]
\vspace*{-4mm}
\centering 
\includegraphics[width=0.9\linewidth,trim=0 36mm 0 0,clip=true]{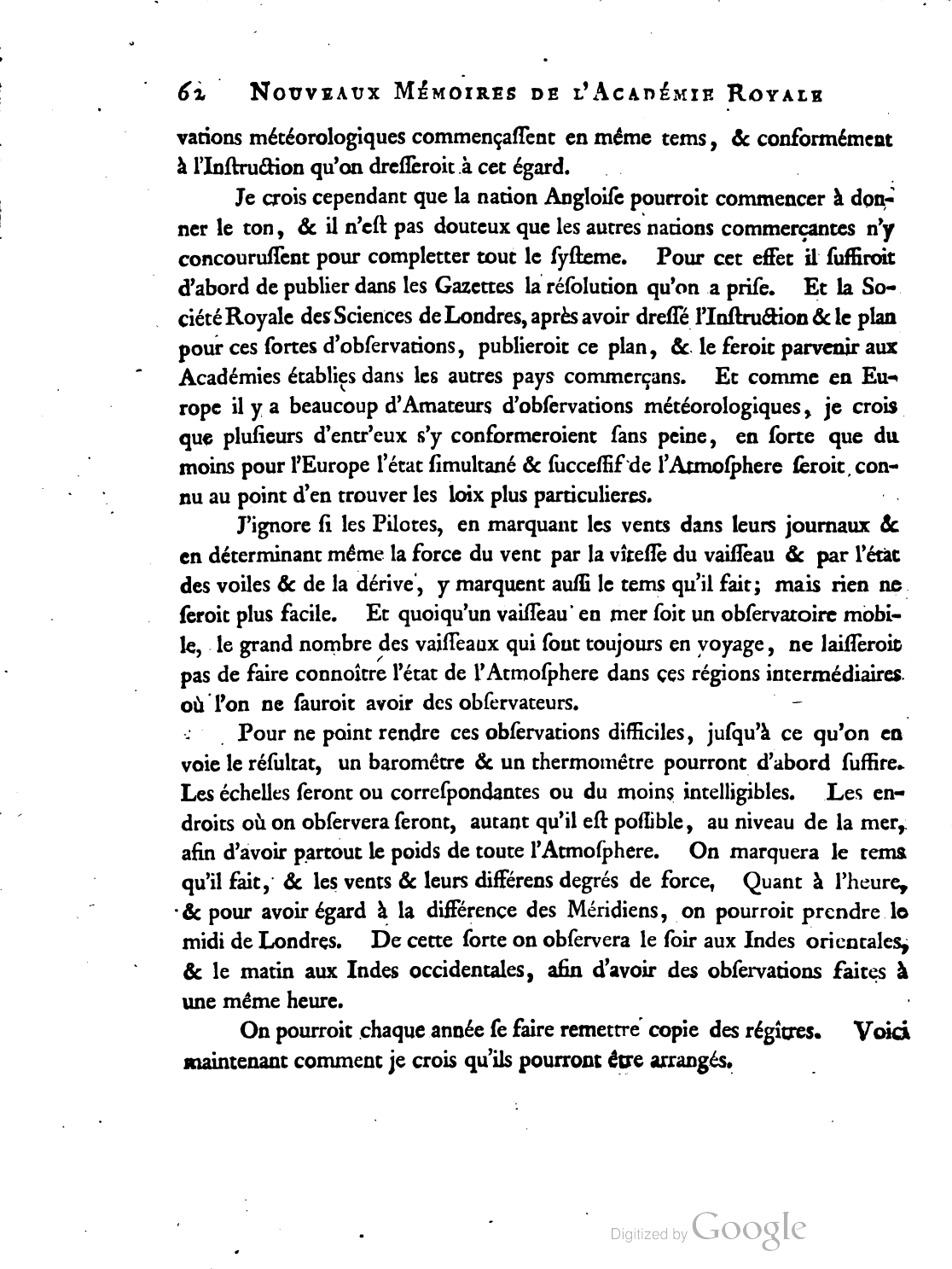}
\label{Lambert_1771_p62}
\end{figure}
\clearpage

\begin{figure}[hbt]
\vspace*{-4mm}
\centering 
\includegraphics[width=0.9\linewidth,trim=0 30mm 0 0,clip=true]{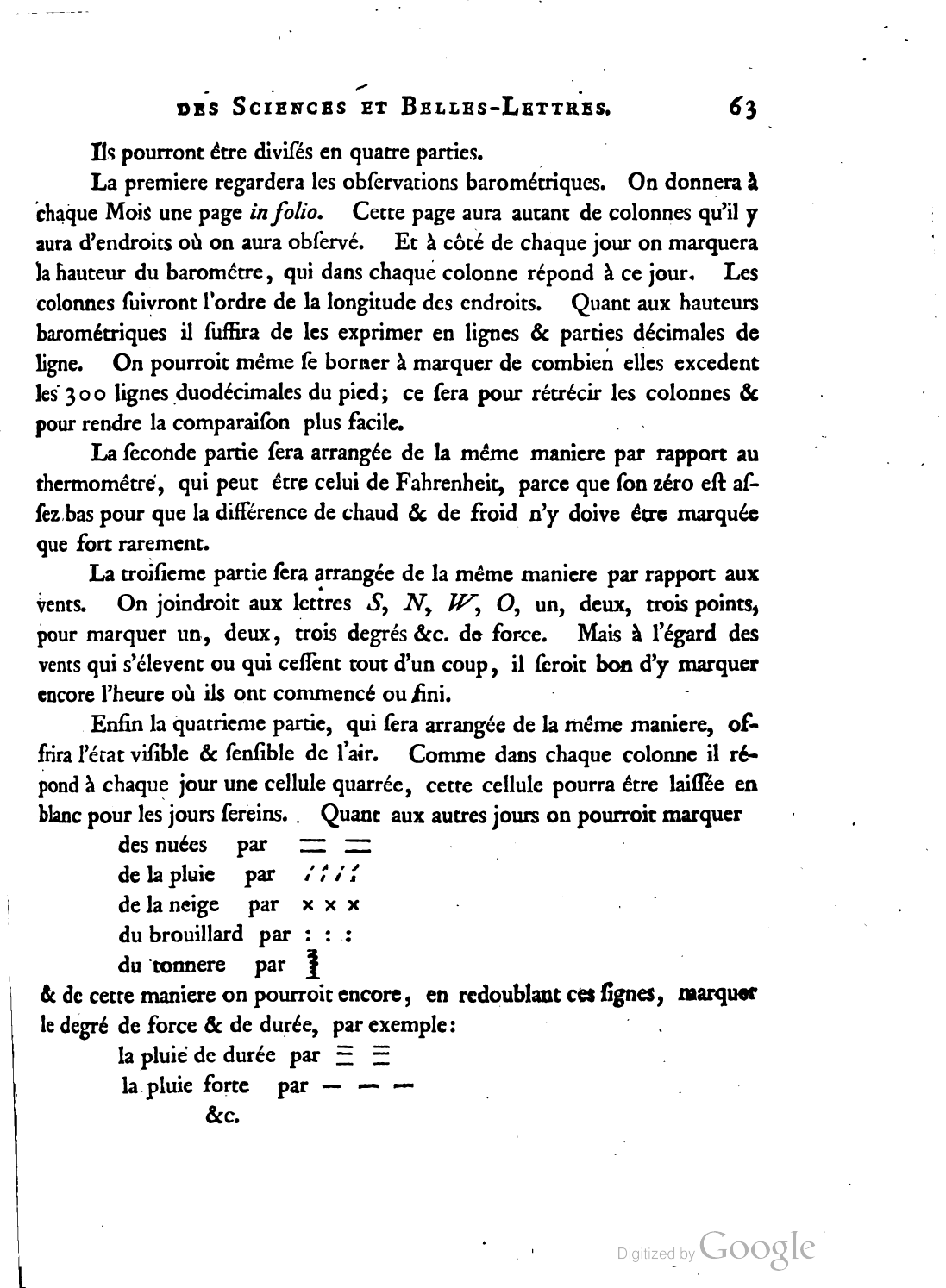}
\label{Lambert_1771_p63}
\end{figure}
\clearpage

\begin{figure}[hbt]
\vspace*{-4mm}
\centering 
\includegraphics[width=0.9\linewidth,trim=0 30mm 0 0,clip=true]{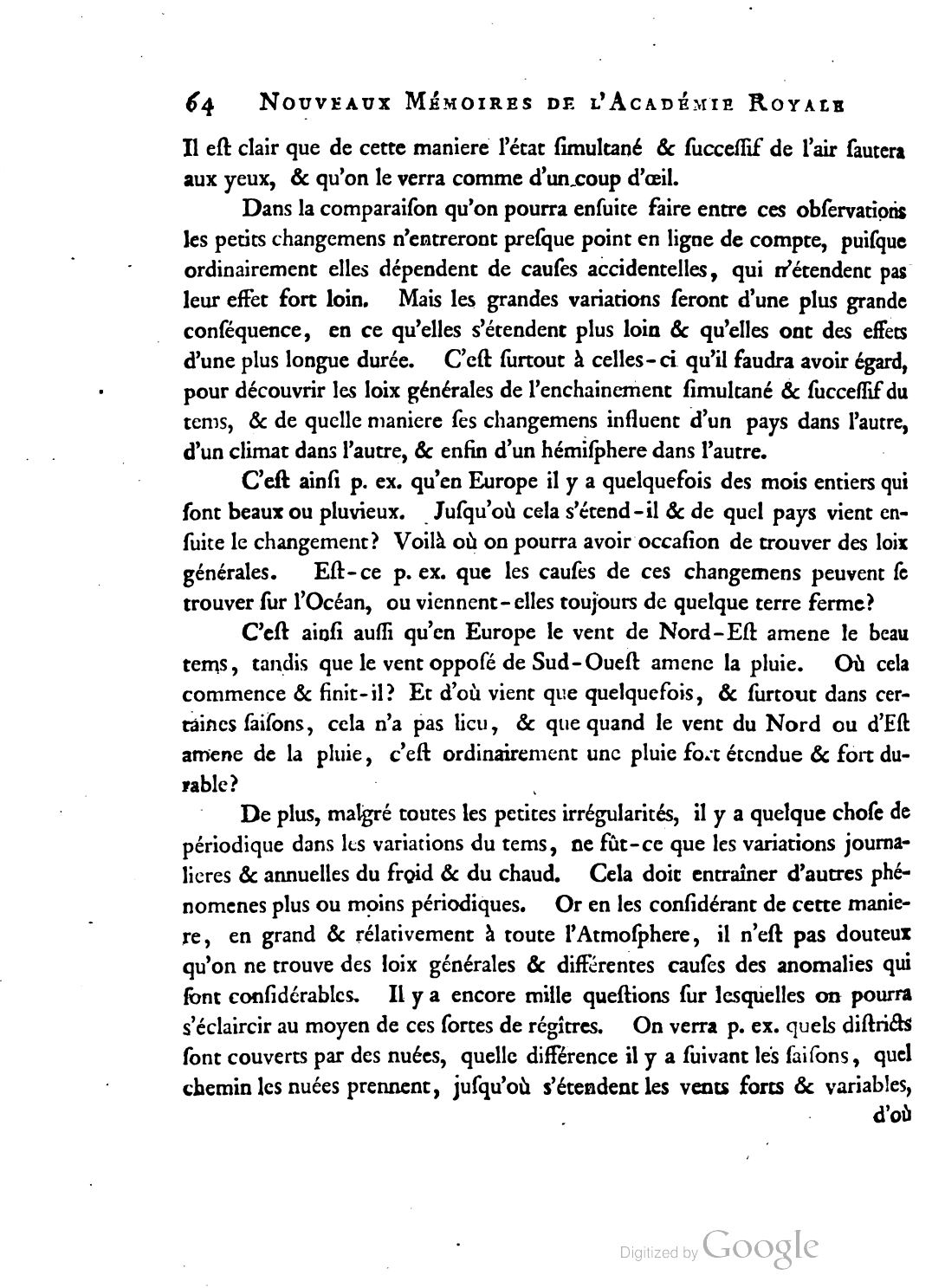}
\label{Lambert_1771_p64}
\end{figure}
\clearpage

\begin{figure}[hbt]
\vspace*{-4mm}
\centering 
\includegraphics[width=0.9\linewidth,trim=0 30mm 0 0,clip=true]{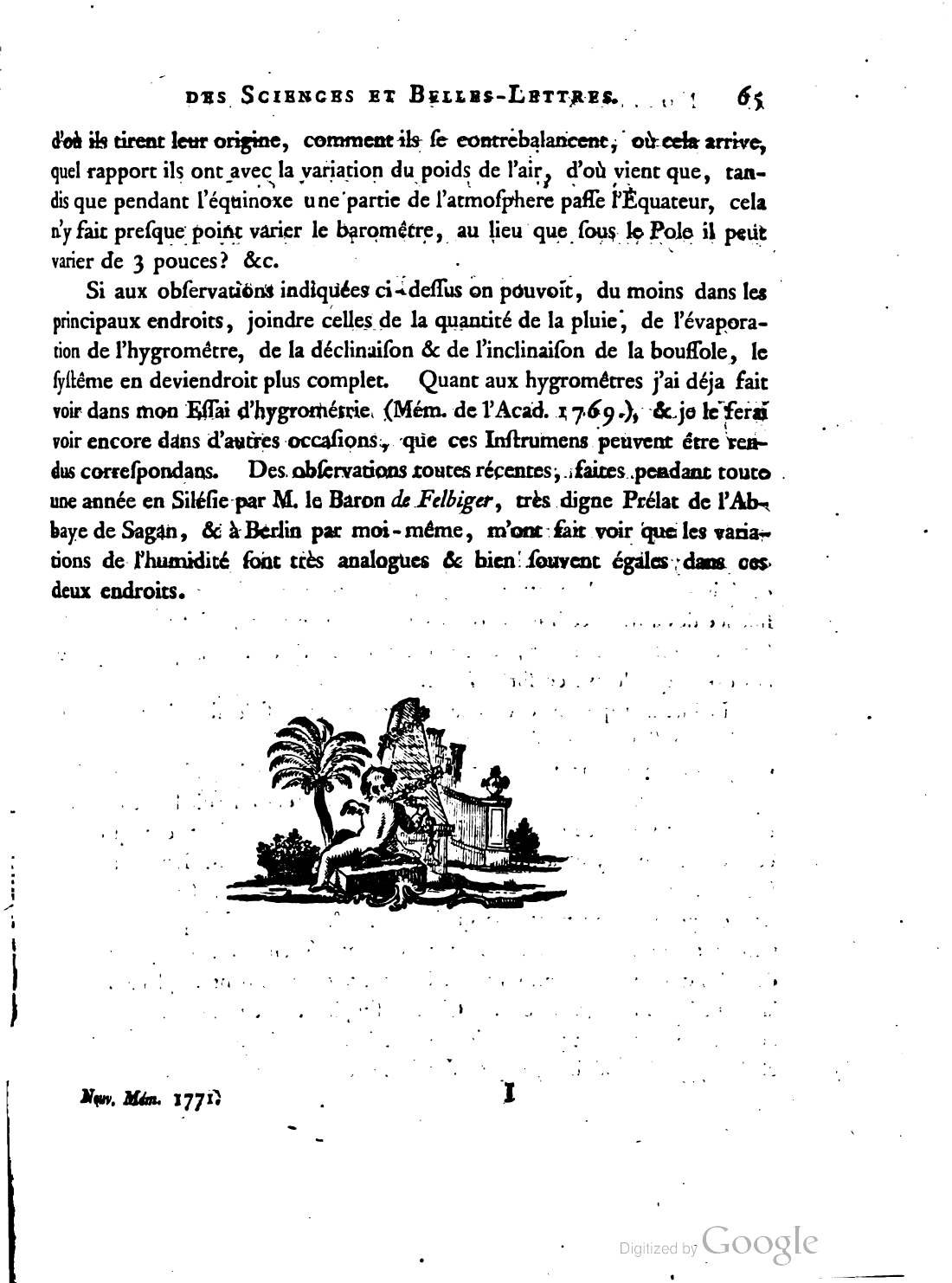}
\label{Lambert_1771_p65}
\end{figure}
\clearpage

\end{document}